\documentclass{article}\hbadness=10000\hfuzz=20pt
\newcommand{\be}{\begin{equation}} 
\newcommand{\ee}{\end{equation}} 

\usepackage{epsfig}
\begin{document}      

\title{{\small Letter}\\
{\bf Satellite measurement of the Hannay angle}}

\author{{\normalsize\bf ALESSANDRO D.A.M. SPALLICCI}\\
{\small UMR 6162, D\'ept. d'Astrophysique Relativiste ARTEMIS, Obs. de la C\^ote d'Azur}\\
{\small BP 4229, Boulevard de l'Observatoire, 06304 Nice, France}\\
{\small Email: spallicci@obs-nice.fr}}

\maketitle
\begin{abstract}
The concept of a measurement of the yet unevaluated Hannay angle, by means of an Earth-bound satellite, adiabatically driven by the Moon, is shown herein.   
Numerical estimates are given for the angles, the orbital displacements,
the shortening of the orbital periods, for different altitudes. It is concluded that the Hannay effect is measurable in high Earth orbits, by means of atomic clocks, accurate Time \& Frequency transfer system and precise positioning. 
\end{abstract}

\begin{flushright}
{Interdisciplinary space science in Giuseppe Colombo's memory}
\end{flushright}

\section{Introduction}

Berry discovered the quantum
geometric phase \cite{ber84} and Hannay \cite{han85} found its classical
correspondent, both contributions giving birth to a lively line of research especially in the mathematical physics 
community. 
Beside the applications to celestial mechanics related to the
non-sphericity and slow rotation of the Earth \cite{celmech}, Berry
and Morgan \cite{bermor96} have attempted to determine the Hannay angle for
the restricted circular three-body problem and provided estimates for the
Earth and Mars revolving around the Sun, under the adiabatic influence of Jupiter, and
for a geostationary satellite, under the influence of the Moon.  
Berry and Morgan incurred in several errors, later amended \cite{spamormet04}
by means of a perturbative approach and a proper Hamiltonian.  \\

Meanwhile, the European Space Agency (ESA) began in 1990 studies on
fundamental physics experiments with clocks on the International Space Station (see \cite{spaletal97} for a review).
The most relevant result was the original conception of the Atomic Clock Ensemble in
Space (ACES) experiment \cite{tfus95}. In an ACES study
\cite{fel99}, the feasibility of a space measurement of the Hannay angle was investigated. \\

The reader is addressed to \cite{spamormet04} and its references for an introduction 
to the Hannay angle, that we briefly summarise herein. 
When an adiabatic system is defined by action-angle variables, 
the Hannay angle is the geometric shift $\theta _H$ in 
the angle variable, arising from a full cycle of the Hamiltonian or in 
other words it is defined if closed curves of constant action return to the
same curves in phase space after a time evolution.
A formal connection between perturbation theory and the Hamiltonian adiabatic approach shows that 
this effect is 
already contained in classical celestial mechanics. 

With reference to a three-body system, A being the attractor, P the perturber and T the test mass, 
the Hannay angle is given by: 

\begin{equation}
  \label{eq:15}
\theta_H=\frac{1261}{16}\pi\left(\frac{M_P}{M_A} \right)^2\left(\frac{R_{T}}{R_{P}} \right)^6  
\end{equation}
where $M$ and $R$ indicate the masses and the orbital radii centred in the A centre of mass.
The Hannay angle manifests itself like a precession and it is observable as forward displacement of the perturbed body on the orbit path at a given radius. 
The Hannay angle of the Earth due to the perturbation by Jupiter \cite{spamormet04} amounts to 
$144 m/y$. 
 
\section{Measurability}

\subsection{Size of the effects}

The Hannay angle is not a new effect but a term in existing equations of motion that has been so far unevaluated separately; being contained in the models of orbital dynamics, it must be dug out of their equations and of the measurements and 
it must be separated from concurring, possibly larger, effects of similar and different time scales. 
In other words, for the topic of measurability, there are three issues to deal with. The first, as we just stated, is to enucleate the Hannay effect from the motion equations that contain it in an embedded form.  Second, an analysis of all perturbations must be carried out sizing their magnitudes and uncertainties. Third, short and long term variables must be analysed timewise.

Furthermore, there are two approximations for an Earth-bound satellite adiabatically driven by the Moon. The first is the non-inertiality of the system revolving around the Sun that can be circumvented by evoking the difference in time scales of the satellite or Moon orbit vis a vis the Earth revolution period. The second is the assumption of circularity of the satellite orbit. \\
The Hannay angle is cumulative and obviously formed continuously. It may be measured at any time including an unfinished perturber revolution, although in the adiabatic approach is defined in closed curves of the perturber (the test mass does not jump forward suddenly).

\begin{table*}
\caption[]{\small Hannay effect for geosynchronous, Galileo, ISS-ACES orbits. 
The Earth, Moon, constant numerical values are: Earth mass $M_A = 5.976 \times 10^{24}$ Kg,   
Earth radius $R_A = 6.37814 \times 10^6 $ m, Moon mass  $M_P = 7.349 \times 10^{22}$ Kg, Moon period $\tau_P =           27.32166$ days, Mean Moon orbital radius $R_P = 3.844 \times 10^8$ m. The universal constants are: 
constant of gravitation G = $6.673 \times  10^{-11}$ m$^3$/(Kg s$^2$), speed of light c = $3 \times  10^{8}$ m/s.} \label{tab1}
\begin{tabular}{lccc} \hline
Parameter & geosynchronous & Galileo-like & ISS-ACES \\ 
\hline
Orbital radius $R_T$ (m) & $4.225\times 10^7$ & $2.656\times 10^7$ & $6.778\times 10^6$\\
Orbital period $\tau_T$ (s)        & 86400                 & 43070.57         & 5552.39 \\
\hline
Forward angle (rad) per Moon orbit 
& $ 6.204 \times 10^{-8}$
& $ 3.829 \times 10^{-9}$  
& $ 1.058 \times 10^{-12}$  \\ 
Forward angle (rad) per satellite orbit 
& $ 2.271 \times 10^{-9}$
& $ 6.983 \times 10^{-11}$  
& $ 2.488 \times 10^{-15}$  \\ 
Forward displacement (m) per Moon orbit 
& $ 2.621 $ 
& $ 1.017 \times 10^{-1}$ 
& $ 7.168 \times 10^{-6}$ \\ 
Forward displacement (m) per satellite orbit 
& $ 9.594 \times 10^{-2}$ 
& $ 1.855 \times 10^{-3}$ 
& $ 1.686 \times 10^{-8}$\\ 
Time shortening (s) per Moon orbit 
& $ 2.331 \times 10^{-2}$  
& $ 1.438 \times 10^{-3}$   
& $ 3.973 \times 10^{-7}$  \\   
Time shortening (s) per satellite orbit  
& $ 3.122 \times 10^{-5}$   
& $ 4.784 \times 10^{-7}$    
& $ 2.198 \times 10^{-12}$  \\   
Velocity increment (m/s) 
& $ 1.11 \times 10^{-6}$ 
& $ 4.308 \times 10^{-8}$ 
& $ 3.037 \times 10^{-12}$ \\
First order Doppler 
& $ 3.701 \times 10^{-15}$ 
& $ 1.436 \times 10^{-16}$ 
& $ 1.012 \times 10^{-20}$ \\
Resolution on diff. posit. along track (m)
& $\ll~3.054 \times 10^{-2}$ 
& $\ll~5.906 \times 10^{-4}$ 
& $\ll~4.657 \times 10^{-9}$ \\ 
Resolution on diff. posit. along track (s)
& $\ll 1.018 \times 10^{-10} $
& $\ll 1.969 \times 10^{-12} $ 
& $\ll 1.552 \times 10^{-17}$ \\ 
\hline
\end{tabular}
\end{table*}

Numerical estimates, ({\it tab. 1}), are given for the angles, the shortening of the orbital periods, the orbital displacements, the effect in dimensionless units and the first order Doppler terms.
The Hannay angle per test mass orbit is given by: 

\begin{equation}
\theta_{H,T~orbit} = \theta_H \frac{\tau_T}{\tau_P}
\end{equation}
where $\tau_P$ is the orbital period of the perturber.
The correspondent forward displacement is given by:

\begin{equation}
D_{H} = \theta_{H}~R_T~~~~~~~~~~~~~~~~~~~~~~~~~~~~~D_{H,T~orbit} = \theta_{H,T~orbit}~R_T
\end{equation}
The shortening of the orbital period is given by:

\begin{equation}
{\Delta \tau_T} = \frac{\theta _{H}}{2\pi} \tau_T~~~~~~~~~~~~~~~~~~~~~~~~~~~~~    
{\Delta \tau_{T~orbit}} = \frac{\theta _{H,T~orbit}}{2\pi} \tau_T     
\end{equation}
The velocity increment is given by:

\begin{equation}
\Delta v_T  = \frac{2\pi a \Delta \tau_{T~orbit}}{(\tau_T - \Delta \tau_{T~orbit})\tau_T} 
\end{equation}
The Doppler effect of a frequency reference on-board would be given by:

\begin{equation}
\frac{\Delta \nu_T}{\nu_T} = \frac{\Delta v_T}{c} 
\end{equation}

There is the need of discriminating the Hannay effect from a faster orbit simply due to a smaller orbital radius. 
This requires a resolution on the differential, i.e. not absolute, positioning along track (the line of sight) better than the Hannay effect, instead occurring
cross track (orthogonal to the line of sight) and {\it tab. 1} reports the requirement on differential ranging along-track (expressed both in terms of space and time):

\begin{equation}
r_{dralt} = R_T - \frac{G M_A  R_T}{R_T \Delta V^2 + G M_A + 2 \Delta V (G M_A R_T)^{1/2}}    
\end{equation} 

\subsection{Discussion of the results}

Satellite positioning measurements have largely improved, but the values for the Hannay effect hint to a formidable challenge.
An helpful contribution could be represented by an atomic clock on board. Indeed, the peculiar feature of the Hannay effect is the forward displacement of the perturbed body on the orbit path at a given radius, 
i.e shorter period for a given radius or else larger radius at a given period.
The Hannay angle may be {\it conceptually} measured by 
coupling the determination of the orbital period to the orbital radius or the orbital velocity (Doppler shift in case of a frequency reference source) to the orbital period or radius. 

The Hannay angle of a
geostationary satellite perturbed by the Moon amounts to $6.204 \times{}10^{-8}$ radians for a perturber period.
The daily effect is $2.207 \times{}10^{-9}$ radians, $9.594 \times{}10^{-2}$ m of {\it forward} displacement along the orbit, $3.122 \times{}10^{-5}$ s of orbital period shortening. The orbital velocity increment is   
$1.11 \times{}10^{-6}$ m/s and translates into a clock first Doppler shift {\it variation} equal 
to $3.701 \times{}10^{-15}$.\\
The values from {\it tab. 1} suggest that measurements are not possible at Galileo and ISS altitudes with today technology, whereas there are some chances at geosynchronous orbits and they are worth the investigations presented herein. 
The Doppler shift and the requirements on differential ranging along-track 
are within feasibility at geo orbit.
The advantage of using atomic clocks is that the associated measurements (e.g. Doppler shift) appear more feasible as opposed to positioning measurements.\\
An optimised strategy for the Hannay effect detection would look into the time and frequency window(s) where the signal is more easily identified, taking advantage that Hannay effect is always present with its no-Keplerian behaviour. 
Three considerations should ease the experimental set-up: i) the measurement can be performed at any time scale (obviously the same time scale where the Hannay effect is tested); ii) the requirement on along track ranging is differential and not absolute; iii) the Hannay effect is time modulated by the differential rotation of the Moon and the satellite. The modeling of such effect may be transformed into a template for signal processing. \\
Finally, although in the classical two-way frequency transfer system the first order Doppler cancels 
\cite{badetal60,vesetal80}, a proper scheme may retain it \cite{spalbus96}. 
In geosynchronous orbit, the first order Doppler due to Hannay is well within the 
frequency stability of atomic clocks but must be extracted from the larger Doppler shift due 
cruising speed, velocity increments due to perturbations and relativistic effects.        

\subsection{Perturbations on a geosynchronous satellite}

Several perturbations and the gravitational influence of a motionless perturber have  
to be taken into account, especially when they determine a precessional motion \cite{milnobfar87,berfar90}. For the determination of the Hannay effect, it is sufficient the measurement of the {\it instantaneous} orbital radius at two different instants coupled with a Doppler shift.\\
At geo altitude, the Hannay effect maybe masked by an orbit which radius is $3$ cm shorter. 
The perturbations which may impose an orbital radius variation larger than, or comparable to, $3$ cm are : $J_2, J_3, J_4$; tidal terms from the Moon, Sun, Venus and the Earth, solar radiation pressure and thermal emission from the spacecraft. Some of them are accurately known perturbations and may be modeled and subtracted out. Anyhow, the state of the art of ranging technology (e.g. Lageos for the gravitomagnetic effect) 
has an accuracy of few mm and thus is far more accurate than modeling.

\section{Conclusions}
In this letter we have investigated the opportunity to perform a first measurement of the Hannay 
angle, in high Earth orbit, making use of a clock on board.

\section{Acknowledgments} 
Thanks are due to Dr S. Feltham, ACES manager and responsible of the 
study for ESA \cite{fel99}.
Dr Ch. Salomon, Ecole Normale Sup. Paris, ACES Principal Investigator and Prof. Sir M. Berry, Bristol Univ., are acknowledged for various discussions; F. Deleflie (Grasse) for a general comment on measurability of quantities in celestial mechanics.
The European Space Agency is acknowledged for awarding the Senior Research Fellowship G. Colombo to A. Spallicci.

\end{document}